\documentclass[twocolumn, prd,preprintnumbers,superscriptaddress,nofootinbib,showpacs]{revtex4-1}

\usepackage{graphicx,slashed,hyperref,color}
\usepackage[normalem]{ulem}
\usepackage[utf8]{inputenc}
\usepackage{amsfonts,amssymb,amsmath}
\usepackage{mathrsfs}
\usepackage{color}
\usepackage{dcolumn}
\usepackage{soul}
\usepackage{bm}
\hypersetup{
   bookmarks=true,         % show bookmarks bar?
   unicode=true,          % non-Latin characters in Acrobat�s bookmarks
   pdftoolbar=true,        % show Acrobat toolbar?
   pdfmenubar=true,        % show Acrobat menu?
   pdffitwindow=false,     % window fit to page when opened
   pdfstartview={FitH},    % fits the width of the page to the window
   pdftitle={My title},    % title
   pdfauthor={Author},     % author
   pdfsubject={Subject},   % subject of the document
   pdfcreator={Creator},   % creator of the document
   pdfproducer={Producer}, % producer of the document
   pdfkeywords={keyword1} {key2} {key3}, % list of keywords
   pdfnewwindow=true,      % links in new PDF window
   colorlinks=true,       % false: boxed links; true: colored links
   linkcolor=blue,          % color of internal links (change box color with linkbordercolor)
   citecolor=blue,        % color of links to bibliography
   filecolor=cyan,      % color of file links
   urlcolor=blue           % color of external links
}

\def\lsim{\mathrel{\rlap{\lower4pt\hbox{\hskip1pt$\sim$}}
    \raise1pt\hbox{$<$}}}         %less than or approx. symbol
\def\gsim{\mathrel{\rlap{\lower4pt\hbox{\hskip1pt$\sim$}}
    \raise1pt\hbox{$>$}}}         %greater than or approx. symbol

\begin{document}

\preprint{CTPU-PTC-18-23, LDU-19-01}

\title{Clockwork inflation with non-minimal coupling}

\author{Seong Chan Park}
\email{sc.park@yonsei.ac.kr}
\affiliation{Department of Physics and IPAP, Yonsei University, Seoul 03722, Korea}
\author{Chang Sub Shin}
\email{csshin@ibs.re.kr}
\affiliation{Center for Theoretical Physics of the Universe, IBS, Daejeon 34051, Korea}

\date{\today}

\begin{abstract}
 We suggest a clockwork mechanism for a Higgs-like inflation with the non-minimal coupling term $\xi \phi^2 R$. The seemingly unnatural ratio of parameters, $\lambda/\xi^2 \sim 10^{-10}$ of the self quartic coupling of the inflaton, $\lambda$, and the non-minimal coupling, $\xi$,  is understood by 
 exponential suppression of $\lambda$ by the clockwork mechanism, instead of a large non-minimal coupling.  
 The portal interaction between the inflaton and the Standard Model (SM) Higgs doublet is introduced as a source of reheating and the inflaton mass.   Successful realization of inflation requires that the inflaton gets a mass around (sub) GeV scale, which would lead to observable consequences depending on  reheating process and its lifetime.

\end{abstract}

\pacs{
%14.60.Pq	%Neutrino mass and mixing (see also 12.15.Ff Quark and lepton masses and mixing)
%14.80.Bn %Standard-model Higgs bosons
%14.65.Ha	%Top quarks
%95.35.+d	%Dark matter (stellar, interstellar, galactic, and cosmological) 
%97.60.Lf	%Black holes
98.80.Cq	%Particle-theory and field-theory models of the early Universe (including cosmic pancakes, cosmic strings, chaotic phenomena, inflationary universe, etc.)
}
\keywords{dark matter, primordial black hole, Higgs inflation, top quark mass}

\maketitle

%%%%%%%%%%%%%%%%%%%%%%%%%%%%%%%%%%%%%%%%%%%%%%%%
\section{introduction}
\label{sec:introduction}
%%%%%%%%%%%%%%%%%%%%%%%%%%%%%%%%%%%%%%%%%%%%%%%%

Higgs inflation is a successful model of inflation based on the Standard Model (SM) of particle physics~\cite{Bezrukov:2007ep}.  A Jordan frame action for Higgs inflation include non-minimal coupling between the inflaton field, or `Higgs' field,  $\phi$ and the Ricci scalar $R$,
\begin{eqnarray}
S_J=\int d^4 x \sqrt{-g} \left(-\frac{M_P^2 +\xi \phi^2 }{2}R -\frac{1}{2}\partial_\mu \phi \partial^\mu \phi -V(\phi)\right),\nonumber \\
\end{eqnarray}
where $M_P\approx 2.4\times 10^{18}~{\rm GeV}$ is the reduced Planck mass.
The potential $V(\phi) =\lambda \phi^4/4$ satisfies the following condition at a large field limit, $\xi \phi^2 \gg M_P^2$, 
\begin{eqnarray}
\lim_{\phi \to \infty} 
\frac{V(\phi)}{(M_P^2+\xi\phi^2)^2} = \frac{\lambda}{4\xi^2}=\text{constant} >0,
\end{eqnarray}
then the potential in Einstein frame has an asymptotically flat plateau and
 accommodates enough number of e-foldings~\cite{Park:2008hz}. 
To fit the cosmological observation, however, the ratio is requested to be extremely small or fine-tuned as
\begin{eqnarray}
\frac{\lambda}{\xi^2} \approx 4.4 \times 10^{-10},
\label{eq:cond}
\end{eqnarray}
and calls for additional explanation~\cite{Kim:2010fq}. 

Conventionally, the small  value is explained by a large non-minimal coupling $\xi \sim {\cal O}(10^{4})$ with $\lambda\sim {\cal O}(1)$~\cite{Bezrukov:2007ep} but it causes un-wanted low scale cut-off for a $\phi$-gravition interaction at around $M_P/\xi$, which is well below the Planck scale~\cite{Burgess:2010zq, Bezrukov:2010jz}. It is also noticed that the small ratio could be obtained by a small quartic coupling $\lambda \ll 1$ at the inflationary scale due to the renormalization group (RG) running with $d\lambda/d\log \mu<0$, for the SM Higgs field~\cite{DeSimone:2008ei, Bezrukov:2009db}. However, it relies on the largish experimental uncertainty in the top quark mass measurement~\cite{Patrignani:2016xqp}. For updated analysis, see \cite{Hamada:2014iga, Hamada:2014wna} and also \cite{Bezrukov:2014bra,Bezrukov:2014ipa}.

%In the case where the SM Higgs field is identified with the inflaton field, within experimental uncertainty especially in the top quark mass measurements~\cite{Patrignani:2016xqp}, this small ratio may be realized by renormalization running of the couplings~\cite{DeSimone:2008ei, Bezrukov:2009db}. For some updated analysis, see \cite{Hamada:2014iga, Hamada:2014wna} and also see \cite{Bezrukov:2014bra,Bezrukov:2014ipa}.

In this paper, we would suggest an alternative, simple explanation for the small ratio by clockwork mechanism. 
We don't need any unnaturally small or large couplings at a defining scale but still realize a successful inflation.  
Especially, we only introduce a mild value of non-minimal coupling, $\xi\sim{\cal O}(1)$ thus the Planck scale cutoff is maintained.

The main idea of clockwork mechanism was first proposed in order to  generate a trans-plankian period of the pseudo scalar inflaton potential~\cite{Choi:2014rja}, and utilized  in more general cases \cite{Choi:2015fiu,Higaki:2015jag,Kaplan:2015fuy}.  It is also generalized to the fields with different spins, and recognized that the localization of the wave functions in the site space resembles that in the deconstruction of the extra dimensional model \cite{Giudice:2016yja}, although the details are not exactly the same \cite{Craig:2017cda,Giudice:2017suc}.  There are also interesting applications of the mechanism for various phenomenological problems such as dark matter, flavor,  composite Higgs, axion $(g-2)$ of muon and  seesaw mechanism~\cite{Kehagias:2016kzt,Farina:2016tgd,Ahmed:2016viu,Hambye:2016qkf,Coy:2017yex,Ben-Dayan:2017rvr, Park:2017yrn, Hong:2017tel,Patel:2017pct, Kim:2018xsp}. We also note that  other possibilities of inflationary scenarios in the context of e.g. linear potential model, hybrid potential modal and Starobinsky's model were considered in \cite{Kehagias:2016kzt,Im:2017eju}. Discussions on continuum limit and connection with linear dilaton models are in \cite{Giudice:2017fmj,Choi:2017ncj}.

This paper is organized as follows: In the next section, Sec.~\ref{sec:cw}, we first review the basic idea of clockwork mechanism for our purpose  then apply to the Higgs-like inflation model in Sec.~\ref{sec:hi}. Finally we conclude in Sec.~\ref{sec:discussion}. 
For definiteness, in below, we consider ``Higgs-like inflation" driven by a SM singlet scalar and the inflation takes place by the interplay between a positive quartic coupling and the non-minimal coupling.

%%%%%%%%%%%%%%%%%%
\section{clockwork mechanism}
\label{sec:cw}
%%%%%%%%%%%%%%%%%%

\begin{figure}[t]
\begin{center}
\includegraphics[width=0.43\textwidth]{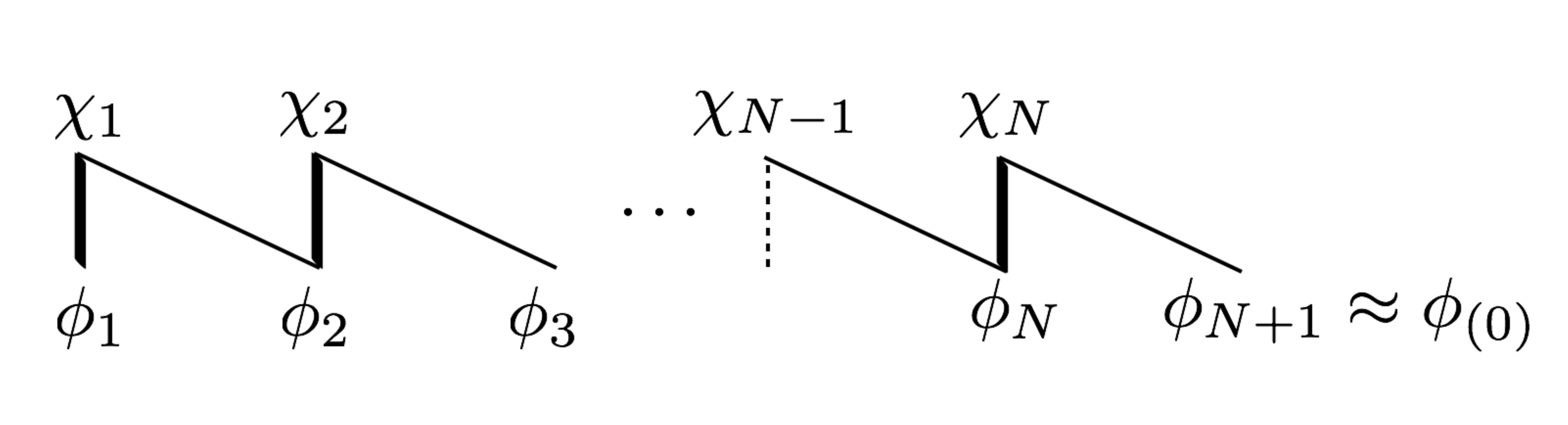} %figures directory 
\end{center}
\caption{A clockwork diagram for a chain of scalar fields.}\label{fig:01}
\end{figure}

A clockwork (CW) mechanism can be described by the  clockwork diagram in Fig.~\ref{fig:01} where a set of heavy fields $\chi_i$ $(i=1,2,\cdots, n)$, and $\phi_i$ $(i=1,2,\cdots, n+1)$ are linked by vertical and diagonal mass parameters, $m_i$ and $M_i$, respectively. The mass parameters are considered to be spurions of symmetries under which the spurions are bi-charged as  $m_i \sim (-Q_{U(1)_{\chi_i}},-Q_{U(1)_{\phi_{i+1}}})$ of $U(1)_{\chi_{i}} \times U(1)_{\phi_{i+1}}$ and $M_i \sim (-Q_{U(1)_{\chi_i}},-Q_{U(1)_{\phi_{i}}})$  of $U(1)_{\chi_{i}}\times U(1)_{\phi_i} $ respectively. 
Under $U(1)_{\phi_i}$,  $\phi_i \sim Q_{\phi_i}$ and $U(1)_{\chi_i}$,  $\chi_i \sim Q_{\chi_i}$. For a scalar potential, such a schematic picture can be discussed more transparently in the context of supersymmetry by constructing following clockwork superpotential:
\begin{eqnarray}
W_{CW} =\sum_{i=1}^N  \left(m_i \chi_i \phi_{i+1} - M_i \chi_i \phi_{i} \right) .
\label{eq:supotcw}
\end{eqnarray}
We  assume that the mass parameters are essentially similar in values so that  $m_i=m$ and $M_i=M$ for $i=1,2,\cdots,N$ below. The ratio is $q=M/m>1$. 
Then the $F$-term scalar potential is calculated as 
\begin{eqnarray}
V_{CW} &=& 	 \sum_{i=1}^{N+1}\left|\frac{\partial W_{CW}}{\partial \phi_i}\right|^2 
	 +\sum_{i=1}^N\left|\frac{\partial W_{CW}}{\partial \chi_i}\right|^2 
\nonumber\\
&=& 
 \sum_{i=1}^{N-1} m^2 \left|\chi_i - q\chi_{i+1}\right|^2 + M^2|\chi_1|^2 + m^2|\chi_N|^2 \nonumber\\
&& +\sum_{i=1}^{N} m^2\left|\phi_{i+1} - q  \phi_{i}\right|^2 
\label{eq:potcw}
\end{eqnarray}
All $\chi_i$s and $N$ combinations of $\phi_i$ are heavy with masses of ${\cal O}(m, M)$. 
To figure out the zero mode of the theory, we can use the Euler-Lagrange equations of motion for $\chi_i$: $\partial W_{CW}/\partial\chi_i=0$ \footnote{ 
	It is noted that the charge assignment itself would not forbid terms composed of a power of $\phi_i^\dagger \phi_i$ in scalar potential and, in principle, they could  additionally contribute to the CW decomposition of the mass eigenstates after supersymmetry is broken. To avoid this complication,  
we would regard the potential in Eq.~\ref{eq:potcw} as our CW model. 
	After integrating out heavy fields $\chi_i$, whose dynamics is essentially irrelevant in our discussion,  we get the effective potential of the form of $V_{CW} = V( (\phi_{i+1}- q\phi_i))$.  
}.  The solutions are iteratively obtained as
\begin{eqnarray}
\phi_k =q^{-(N+1-k)} \phi_{N+1}, ~~(i=1,\cdots, N).
\end{eqnarray}
In order to find the zero mode scalar field, we  see the kinetic terms after inserting the solution of  the equations of motion:
\begin{eqnarray}
{\cal L}_{\rm kin}&=& 
-\frac{1}{2} \sum_{i=1}^{N+1} |\partial_\mu \phi_i|^2 \nonumber \\
&\equiv& -\frac{1}{2}|\partial_\mu \varphi_0|^2,
\end{eqnarray}
where $\varphi_0$ is the canonically normalized CW zero mode scalar field as
\begin{eqnarray}
\varphi_{0} =\sqrt{\frac{1-q^{-2(N+1)}}{1-q^{-2}}} \phi_{N+1} 
\equiv \sqrt{{\cal N}_2(q)}\phi_{N+1} 
\label{eq:CWzero}
\end{eqnarray}
where the conveniently defined numerical factor, ${\cal N}_2$.
Approximately, a gear field $\phi_k$, $k \in (1,N+1)$, whose value is determined by equations of motion, is related with the zero mode as
\begin{eqnarray}\label{gears_sol}
\phi_k \approx \frac{\sqrt{1-q^{-2}}}{q^{N+1-k}}\varphi_0.
\end{eqnarray}
One can notice that the zero mode is close to $\phi_{N+1}$ (indeed $\phi_{N+1}=\varphi_{0}$ when $q\to \infty$) as we have  depicted in Fig.~\ref{fig:01} and is said to be `localized at $i=N+1$ site'.  The other end point is for $\phi_1\sim q^{-N}\varphi_{0}$ so that the effective coupling of the zero mode to the other sector of the model, which is described by an operator of dimension $n$, $\hat{{\cal O}}_n$,  is highly suppressed as
\begin{eqnarray}
\sigma\hat{{\cal O}}_n\phi_1^{p} \sim \sigma q^{-pN} \hat{\cal O}_n\varphi_{0}^{p}
\end{eqnarray}
with a positive power, $p>0$. The effective coupling is now read to be $\sigma_{\rm eff} \sim  \sigma q^{-pN}\ll \sigma $ and its size is naturally small with $q^N \gg 1$. This explains how the CW mechanism would address hierarchy problems for seemingly unnatural small parameters. 
We would suggest a model of inflation based on the CW mechanism in the next section.

%%%%%%%%%%%%%%%%%%%%%%%%%%%%%%%%%%%%
%%%%%%%%%%%%%%%%%%%%%%%%%%%%%%%%%%%%

%%%%%%%%%%%%%%%%%%%%%%%%%%%%%%
\section{Clockwork Higgs-like inflation}
\label{sec:hi}
%%%%%%%%%%%%%%%%%%%%%%%%%%%%%%

%%%%%%%%%%%%%%%%%%%%%%%%%%%%%%
\subsection{Clockwork mechanism for inflation}
%%%%%%%%%%%%%%%%%%%%%%%%%%%%%%

The action for `clockwork Higgs-like inflation' is introduced with non-minimal coupling terms $K(\phi_i)R$ and the potential terms $V(\phi_i)$ with the CW potential $V_{CW}$: 
%\begin{widetext}
\begin{eqnarray}
S_J &=&\int d^4 x \sqrt{-g} \left[\frac{M_P^2 +K}{2}R  \right. 
\nonumber  \\
&& \left.\hskip 1.4cm -\sum_i \frac{\left(\partial_\mu \phi_i\right)^2}{2}  -V_{CW}  - V_{\rm inf}\right], 
\end{eqnarray}
%\end{widetext}
%
where the non-minimal coupling term and the CW potential are given as 
\begin{eqnarray}
K(\phi_i) 
&\equiv& \sum_{i=1}^{N+1} \xi_i \phi_i^2,\nonumber\\
V_{CW}(\phi_i) &=& \sum_{i=1}^N \frac{m^2}{2}(\phi_{i+1} - q \phi_i)^2
\end{eqnarray}
with positive $\xi_i ={\cal O}(1)$ and $q>1$. Here we consider the 
CW gears as real scalar fields. 
The quartic potential, which is responsible for inflation, is introduced only for $\phi_{1}$ as 
\begin{eqnarray} 
V_{\rm inf}(\phi_{1}) = \frac{\lambda_\phi}{4}\phi_{1}^4, 
\end{eqnarray} which breaks the CW shift symmetry.\footnote{In fact, the non-minimal coupling term also breaks the CW shift symmetry, which might cause the set-up radiatively unstable especially when $\xi$ is sizable. The loop corrections to the zero mode scalar potential mediated by heavy modes  can be estimated as 
\begin{eqnarray}\label{loop_corr}
\Delta V = \frac{ c_1 \xi^2 }{16\pi^2} \frac{m^4}{M_P^2} \varphi_0^2  
 + \frac{c_2 \xi^4 }{16\pi^2}\frac{m^4}{M_P^4} \varphi_0^4 +\cdots. 
\end{eqnarray}
One could think the couplings and masses as the renormalized values as 
$\lambda =\lambda_{\rm bare} +\delta\lambda_{\rm loop}$ and $m = m_{\rm bare} + 
\delta m_{\rm loop}$ to maintain the CW symmetry.  However, 
it is also identified that those corrections in (\ref{loop_corr}) are subdominant for $m\lesssim 10^{-2} M_P/\xi$ without tuning.} %and more focus on the phenomenological implications. }

 Taking the masses of the CW heavy modes greater than the inflation scale (i.e., $m\gg  V_{\rm inf}^{1/2}/M_P\sim H_{\rm inf}$), we can safely integrate out the heavy fields, and get the effective action for the CW zero mode. We will come back to the effect of heavy modes later. 
In the Einstein frame, 
\begin{eqnarray}
S &=& \int d^4 x\sqrt{-g} \left[\frac{M_P^2}{2}R  \
- \frac{3M_P^2(\partial_\mu K)^2}{4(M_P^2+K)^2}\right. \nonumber\\
&&\left. \quad- \frac{\sum_i (\partial_\mu \phi_i)^2}{2(1+K/M_P^2)} -\frac{V_{CW}+V_{\phi_1}}{(1+K/M_P^2)^2}\right] \nonumber\\
&=& \int d^4 x \sqrt{-g}\left[\frac{M_P^2}{2} R 
- \frac{Z}{2} (\partial_\mu\varphi_{0})^2 
- U  \right].
\end{eqnarray}
The last line is obtained by inserting the clockwork solution for the gear fields (\ref{gears_sol}), which yields $Z$ and $U$ as  
\begin{eqnarray}
Z(\varphi_{0}) &=& \frac{M_P^2(1+ \xi \varphi_{0}^2  + 6\xi^2\varphi_{0}^2)}{(M_P^2+ \xi\varphi_{0}^2)^2},  \nonumber\\
U(\varphi_{0}) &=& \frac{\lambda \varphi_0^4}{
4( 1 + \xi \varphi_{0}^2/M_P^2 )^2}, 
\end{eqnarray} and 
\begin{eqnarray} 
\xi=\frac{\sum_{i=1}^{N+1} \xi_{i} q^{-2(N+1-i) }}{{\cal N}_2(q)}, \ 
\lambda = \frac{ \lambda_\phi q^{-4N} }{{\cal N}_2 (q)^2}, 
\end{eqnarray}
where ${\cal N}_2(q)$ is defined in (\ref{eq:CWzero}).	
Because the non-minimal coupling term is universally contributed by
all gear fields with $\xi_i \sim \xi_{N+1}$,  the effective coupling is
not suppressed as $\xi={\cal O}(1)$, while the quartic coupling is 
dominated by the first gear field, $\phi_1$. Therefore $\lambda \sim \lambda_\phi q^{-4N}$.

Having the effective theory for the zero mode fields, 
we get the effective coupling
\begin{eqnarray}
\frac{\lambda}{\xi^2} \approx \frac{\lambda_\phi}{\xi^2} \frac{1}{q^{4N}} \ll 1,
\end{eqnarray}
which explains the small value $10^{-10}$ taking 
$q^{4N} \sim 10^{10}$ with $\lambda_\phi\sim\xi={\cal O}(1)$.\footnote{Higgs-like inflation with a very small quartic coupling was also discussed in different contexts such as \cite{Kaiser:1994vs,Komatsu:1999mt,Tenkanen:2016twd,Alanne:2016mpa} and many others.}

 The field value of $\varphi_0$ during inflation for the CMB scale, $(\varphi_0)_*$, is determined by the required e-folding number $N_e$ as  
\begin{eqnarray}
N_e \simeq \frac{\xi (\varphi_0)_{*}^2}{M_P^2}  = 50-60.
\end{eqnarray}
The initial value of $\varphi_0$ is of the similar size of $M_P$, $(\varphi_0)_*\sim (8/\sqrt{\xi}) M_P$, so 
  we would carefully check if the heavy fields would spoil the inflation dynamics because of the mixing from the quartic potential. 
 Let us discuss it with two field decomposition of the scalar fields: $\varphi_0$ and $\varphi_1$ as the eigenstates of the clockwork potential, where $\varphi_1$ represents a heavy mode, which would potentially affect the inflationary dynamics closely. Then, 
\begin{eqnarray}
K(\phi_i) &=& \xi \varphi_0^2 + \xi_1 \varphi_0 \varphi_1 + \xi_2 \varphi_1^2, \nonumber\\
V_{\rm inf}(\phi_{1}) &=& \lambda_\phi q^{-4 N} \varphi_0^4 
+ 3 \lambda_\phi q^{-3N} \varphi_0^3 \varphi_1 + \cdots.
\end{eqnarray}  
In the scalar potential, $(V_{CW} + V_{\rm inf})/(1+K/M_P^2)^2$, the dominant tadpole contribution for the heavy modes
is coming from the quartic potential ($\sim m^2\varphi_1^2 + \lambda_\phi q^{-3N}\varphi_0^3 \varphi_1$), which gives the shift of the heavy field as
\begin{eqnarray}
\langle \varphi_1\rangle \sim \lambda_\phi q^{-3 N} \varphi_0^3/ m^2, 
\end{eqnarray} for  $m^2 \gg \lambda_\phi q^{-2 N} \varphi_0^2 \simeq 10^{-5} \sqrt{\lambda_\phi} N_e M_P^2$ during inflation. 
The CW heavy modes are still heavier than the inflaton scale, so we can integrate them out and get the 
effective potential of the zero mode. 
For the large field value of $\varphi_0$ (during inflation), the effective potential is corrected as 
\begin{eqnarray}
U_{\rm eff}(\varphi_0) = \frac{\lambda_\phi q^{-4 N}}{\xi^2} 
\left[1 - \frac{ 2M_P^2}{\xi\varphi_0^2}  + {\cal O}\left(\frac{\lambda_\phi q^{-2 N}\varphi_0^2}{m^2}\right)\right].
\end{eqnarray}
For the initial value of $\varphi_0$, 
$(\varphi_0)_*\sim \sqrt{N_e/\xi} M_P$, the heavy field contributions 
for the inflation dynamics are suppressed as $q^{-2 N} N_e^2/\xi \ll 1$,   
compared to the leading contribution to the slow roll parameters.   
 In short, our treatment of inflaton potential is robust and  the corrections from the heavy gear fields are small.

%%%%%%%%%%%%%%%%%%%%%
\subsection{Higgs portal with Clockwork}
%%%%%%%%%%%%%%%%%%%%%

It is an intriguing possibility that the standard model Higgs doublet field, $H$,  has a portal interaction with other sector of scalar field(s), $\lambda_{H\phi_i} |H|^2 \phi_i^2$. This is particularly interesting because the current measurement of top quark mass may imply metastable electroweak vacuum \cite{Isidori:2001bm, Degrassi:2012ry, Alekhin:2012py} (also see \cite{Chigusa:2017dux, Andreassen:2017rzq} for the state-of-the-art calculation of the decay rate) and the Higgs portal interactions would remedy the problem~\cite{Ko:2014eia}. From the RG equation of $\lambda_H$, $\lambda_{H\phi}$, and $\lambda_\phi$, we can obtain the positive values of $\lambda_H$ and $\lambda_\phi$ for all scales~\cite{Falkowski:2015iwa}. 
 In our set-up, we introduce a coupling only between the Higgs and 
the first gear field, $\phi_{1}$ \cite{Kim:2017mtc}, in order not to disturb the inflation dynamics through the radiative corrections from the Higgs loop, but still yield the meaningful coupling between the Higgs and the inflaton field \cite{Lerner:2009xg}.

Now the scalar potential is extended for the SM Higgs and the singlet fields,  \begin{eqnarray}
V &=&  V_{\rm CW} +V_{\rm inf} + V_{H\phi}, \nonumber \\
V_{H\phi} &=& \mu_H^2 |H|^2 + \lambda_H |H|^4 + \lambda_{H\phi} |H|^2 \phi_{1}^2,  \end{eqnarray}
and the non-minimal coupling is also extended as
$K(H, \phi_i) =  \xi_H |H|^2  + K(\phi_i)$.

It is known that during the inflation, the Higgs gets a effective mass squared from the non-minimal coupling as $-\xi_H H_{\rm inf}^2 \sim - \xi_H q^{-4N} M_P^2$, therefore the term makes the Higgs unstable for a positivie $\xi_H$ \cite{Espinosa:2007qp}. However from the Higgs portal coupling, there is the another source that makes the Higgs stable during inflation from the Higgs portal interactions \cite{Lebedev:2012sy} as
\begin{eqnarray}
\Delta m_H^2 = \lambda_{H\phi}q^{-2N}\langle\varphi_0^2\rangle \simeq \lambda_{H\phi} q^{-2N} M_P^2/\xi 
\end{eqnarray}
which is much bigger than the contribution from the Higgs non-minimal coupling. 
For the positive $\lambda_{H\phi}$, the Higgs can be stable during inflation even if the the Higgs quartic is negative at high energy scale. 

After the end of inflation, the inflaton   will start to oscillate, and 
the Higgs particles could be produced through parametric resonance in the preheating stage \cite{Kofman:1994rk,Kofman:1997yn}. There are several studies about the bound on $\lambda_{H\phi}$ in order not to destabilize the Higgs field after inflation with the assumption that 
the inflaton field  is oscillating with a quadratic potential around its minimum \cite{Herranen:2015ima,Ema:2016kpf,Kohri:2016wof,Enqvist:2016mqj, Ema:2017loe}. 
 For   the Higgs-like inflation with a large non-minimal coupling constant ($\xi\gg 1$), such a 
 quadratic approximation is valid for a long time until  the oscillating amplitude 
 of the
 canonically normalized inflaton field becomes of ${\cal O}(M_P/\xi)
= 10^{-4}M_P \ll M_P$, so that most of preheating history is the same as that of the literatures.
In our case the situation is a little bit different, because 
  $\xi \sim O(1)$ and
there is no source of constant mass terms for $\varphi_0$ except the Higgs VEV. It starts to roll 
dominantely with a quartic potential, $U(\phi_0)\simeq \lambda_\phi q^{-4 N} \varphi_0^4$, which means that we cannot simply take the quadratic approximation for the 
motion of $\varphi_0$. 
 If thermalization arises much quicker than the case with a quadratic potential, the Higgs could be trapped at the origin due to its thermal potential.
Therefore, 
it needs further studies for the Higgs stability with $\lambda_{H\phi}$ after inflation.  On one hand,
if $\lambda_{H\phi}$ is negative the Higgs is destabilized during inflation, and spoils the previous discussion. In this sense, we  naturally take $\lambda_{H\phi}>0$.

At present Universe, the clockwork gears are very heavy so that we cannot produce it. 
For the zero mode, the Higgs portal provides the mass term as 
\begin{eqnarray}
V_{\rm eff} =  (q^{- 2N}  \lambda_{H\phi}  |H|^2) \varphi_0^2 + (\lambda_\phi q^{-4 N}) \varphi_0^4.
\end{eqnarray}
The mass is $m_{\varphi_0} \sim q^{-N} v$ and the couplings between the Higgs particles and the zero mode particles are 
\begin{eqnarray}
{\cal L} = (\lambda_{H\phi} q^{- N} v)^2 \delta\varphi_0^2 +  \lambda_{H\phi} q^{-2 N} v\,  h \delta\varphi_0^2  + \cdots 
\end{eqnarray}
The Higgs can decay as $h\to \delta\varphi_0\delta\varphi_0$ with the coupling $q^{-2N} v 
\sim 10^{-5} v$, which is compatible with  the current LHC bound on the Higgs invisible decay, ${\rm Br}(h\to {\rm invisibles}) \lsim 24\%$~\cite{Tanabashi:2018oca}. 
The numerical value of $q^{-N}$ leads to a light mass for $\varphi_0$:
\begin{eqnarray} 
m_{\varphi_0} \sim 10^{-2.5} v \sim {\cal O}(0.1-1)\rm GeV. 
\end{eqnarray}
Our inflationary scenario thus predicts the GeV scale light particles which are coupled to the Higgs weakly, whose experimental search would be extremely interesting and deserves further study~\cite{scp}.

%%%%%%%%%%%%%%%%%%%%%%%%%%%
\section{Conclusion}
\label{sec:discussion}
%%%%%%%%%%%%%%%%%%%%%%%%%%%

Higgs(-like) inflation is an attractive model of inflation which explains the cosmological data with the collaborative helps from the non-minimal coupling and the inflaton potential.  On the other hand,  the required ratio of the self-coupling constant ($\lambda$) and the non-minimal coupling ($\xi$) is unnaturally small,  $\lambda/\xi^2 \sim 10^{-10}$, which thus requires additional explanation. Clockwork mechanism  provides  an interesting answer. By construction, the effective coupling of the inflaton potential  $\lambda/\xi^2 =(\lambda_\phi/\xi^2) q^{-4N}$ is efficiently suppressed by the factor $q^{4N} \sim 10^{10}$.

The constructed clockwork framework leads to interesting implications to the rest of cosmological history and observational consequences:
%Several discussions on consistency, reheating, tensor-to-scalar ratio($r$), and late time implication are in order. 
\begin{itemize}
\item{During inflation:}
Having $\xi\sim {\cal O}(1)$ and $\lambda\sim{\cal O}(1)$ in our setup, the unitarity problem of conventional Higgs inflation~\cite{Burgess:2010zq, Bezrukov:2010jz} would be relieved. The stochastic quantum fluctuations of the scalar fields coupled to the inflaton are quite suppressed because they are all heavy ($M\gg H_{\rm inf}$) and their effects on the inflaton potential is subleading. The SM Higgs also can be stable thanks to the large positive mass squared from the Higgs-inflaton coupling.

\item{Reheating:}
Just after the inflation, the dominant potential of the inflaton 
	is quartic, $U\simeq \lambda_\phi q^{-4N}\varphi_0^4$.  
Because all other CW gear fields are heavy enough, 
we can safely focus on  dynamics of the inflaton and the Higgs fields with quartic potentials 
$U\sim \lambda_\phi q^{-4N}\varphi_0^4 + \lambda_{H\phi} q^{-2N}\varphi_0^2 |H|^2 
	 +\lambda_H |H|^4$ and the initial conditions as  $\langle\varphi_0\rangle\sim M_P/\sqrt{\xi}$, and $\langle H\rangle \ll \langle \varphi_0\rangle$. 
	 Since we cannot take a quadratic approximation for the potential of $\varphi_0$, 
the dynamics for the production of the Higgs and other SM particles are all involved. More detailed study about the (p)reheating in this kind of system (with a scale invariant scalar potential for the Higgs-inflaton) would be quite interesting~\cite{scp}.

\item{Tensor-to-scalar ratio:} Even though $N \approx 10/\log q$ number of fields are involved in CW framework, effectively single field plays the role of inflaton so that a small tensor-to-ratio, $r\sim 10^{-3}$ is expected~\cite{Park:2008hz} even though a largish $r\sim 0.1$ is not completely ruled out~\cite{Hamada:2014iga, Hamada:2014wna}.

 \item{Late time dynamcis of the inflaton:} If the clockwork mechanism for the inflation works, the inflaton mass at its minimum ($\varphi_0=0$) is given by the Higgs portal interaction, and around (sub) GeV. In our minimal example, it has a $Z_2$ symmetry, so the inflaton  (i.e., the quanta of the inflaton field) could contribute to the measured amount of dark matter $\Omega_{DM} h^2\simeq 0.12$~\cite{Aghanim:2018eyx}.  The relic density of the inflaton depends on the reheating procedure, and we could give further constraints on the the size of the coupling $\lambda_{H\phi}$ or the breaking scale of $Z_2$, which can predict the observations in experiments searching for ALPs. 
We remain it as a future work. 

\end{itemize}

\vspace{.5cm}

%%%%%%%%%%%%%%%%%%%%%%%%%%%%%%%%%%%%%%%%%%%%%%%%
%%%%%%%%%%%%%%%%%%%%%%%%%%%%%%%%%%%%%%%%%%%%%%%%
\acknowledgments
This work is supported by the National Research Foundation of Korea (NRF) grant funded by the Korean government (MSIP) (No. 2016R1A2B2016112) and (NRF-2018R1A4A1025334) (SCP), and 
by IBS under the project code, IBS-R018-D1 (CSS).

%%%%%%%%%%%%%%%%%%%%%%%%%%%%%%%%%%%%%%%%%%%%%%%%
%%%%%%%%%%%%%%%%%%%%%%%%%%%%%%%%%%%%%%%%%%%%%%%%

\bibliographystyle{apsrev4-1}
\bibliography{cwhi}

\end{document}